%% file: sample-manuscript.tex
\documentclass[sigconf,authorversion]{acmart}
\AtBeginDocument{%
  }

\setcopyright{cc}
\setcctype{by}
\acmJournal{PACMHCI}
\acmYear{2026} \acmVolume{10} \acmNumber{3} \acmArticle{ETRA024}
\acmMonth{5} \acmDOI{10.1145/3806038}

\newcommand{\revision}[1]{\textcolor{black}{#1}}

\usepackage{fvextra}

\DefineVerbatimEnvironment{PromptCode}{Verbatim}{
  breaklines=true,
  breakanywhere=true,
  breaksymbolleft={},
  breaksymbolright={},
  fontsize=\scriptsize
}

\newenvironment{PromptText}
  {\begin{quote}\footnotesize\raggedright}
  {\end{quote}}

\begin{document}

\title{Lazy or Efficient? Towards Accessible Eye-Tracking Event Detection Using LLMs}


\author{Dongyang Guo}
\affiliation{%
	\institution{Technical University of Munich}
    \city{Munich}
    \country{Germany}
	\streetaddress{Marsstraße 20-22}
	\postcode{80335}
}
\affiliation{%
  \institution{Munich Center for Machine Learning (MCML)}
  \city{Munich}
  \country{Germany}
}
\email{dongyang.guo@tum.de}
\orcid{0009-0009-2088-9330}

\author{Yasmeen Abdrabou}
\orcid{0000-0002-8895-4997}
\affiliation{%
	\institution{Technical University of Munich}
    \city{Munich}
    \country{Germany}
	\streetaddress{Marsstraße 20-22}
	\postcode{80335}
}
\affiliation{%
  \institution{Munich Center for Machine Learning (MCML)}
  \city{Munich}
  \country{Germany}
}
\email{yasmeen.abdrabou@tum.de}

\author{Enkelejda Kasneci}
\orcid{0000-0003-3146-4484}
\affiliation{%
	\institution{Technical University of Munich}
    \city{Munich}
    \country{Germany}
	\streetaddress{Marsstraße 20-22}
	\postcode{80335}
}
\affiliation{%
  \institution{Munich Center for Machine Learning (MCML)}
  \country{Germany}
}
\email{enkelejda.kasneci@tum.de}

\renewcommand{\shortauthors}{Guo et al.}

\begin{teaserfigure}
    \centering
    \includegraphics[width=1.0\linewidth,height=0.4\textheight,
        keepaspectratio
    ]{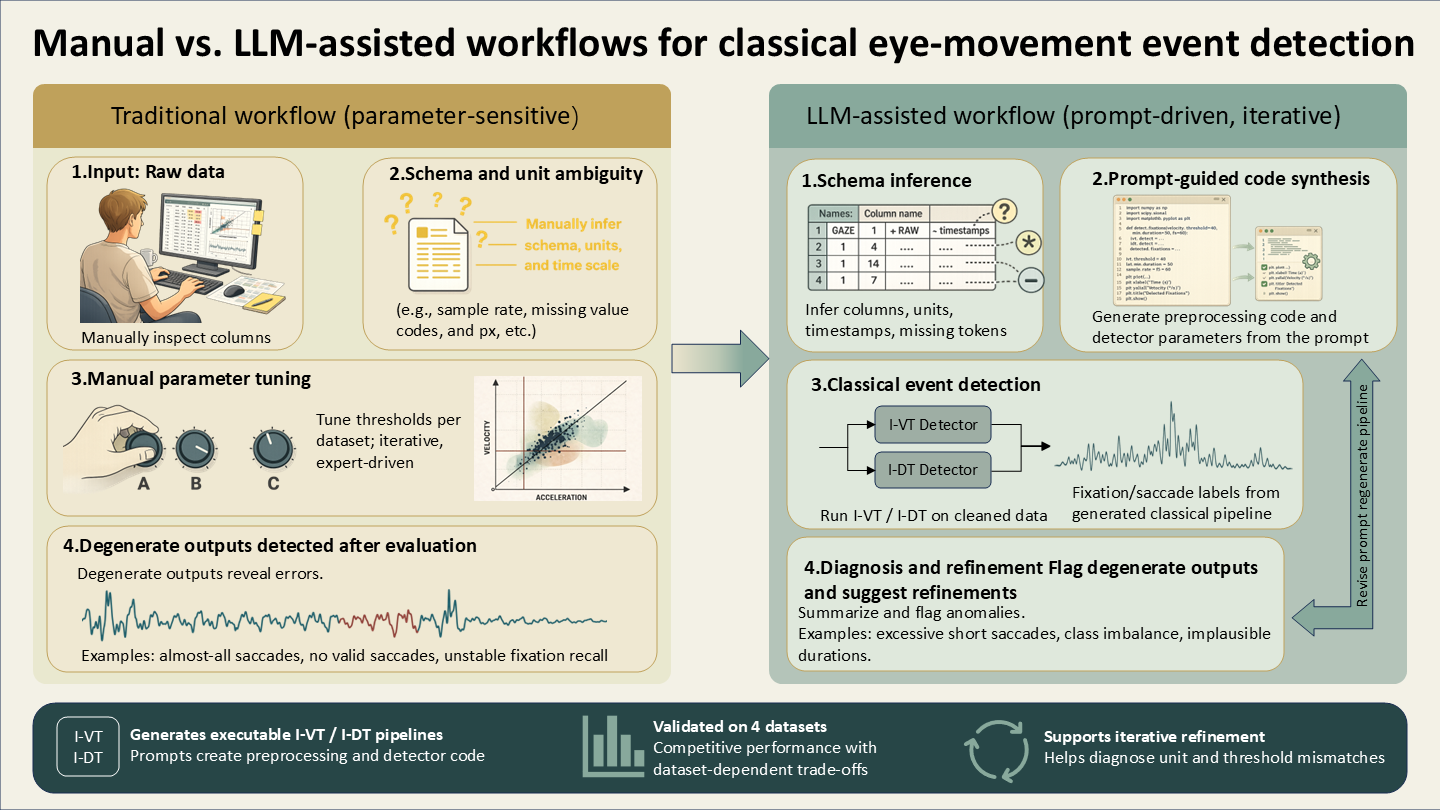}
\caption{This figure compares our method with the traditional manual approach. In particular, the LLM-assisted workflow on the right illustrates our pipeline: (1) starting from raw data, the LLM automatically identifies relevant fields, assesses data quality characteristics, and generates an easy-to-understand summary; (2) users specify their requirements in natural language, and the LLM produces executable data-cleaning code together with parameterized implementations of selected event detectors (e.g., I-VT, I-DT); (3) the cleaned data is then passed to a detector set composed of multiple event-labeling algorithms; and (4) the results and explanatory reports are returned, allowing users to iteratively refine the code and parameters by editing prompts and rerunning the detection process.}
\label{fig:1}
\end{teaserfigure}


\begin{abstract}
Gaze event detection is fundamental to vision science, human-computer interaction, and applied analytics. However, current workflows often require specialized programming knowledge and careful handling of heterogeneous raw data formats. Classical detectors such as I-VT and I-DT are effective but highly sensitive to preprocessing and parameterization, limiting their usability outside specialized laboratories. This work introduces a code-free, large language model (LLM)-driven pipeline that converts natural language instructions into an end-to-end analysis. The system (1) inspects raw eye-tracking files to infer structure and metadata, (2) generates executable routines for data cleaning and detector implementation from concise user prompts,  (3) applies the generated detector to label fixations and saccades, and (4) returns results and explanatory reports, and allows users to iteratively optimize their code by editing the prompt. Evaluated on public benchmarks, the approach achieves accuracy comparable to traditional methods while substantially reducing technical overhead. The framework lowers barriers to entry for eye-tracking research, providing a flexible and accessible alternative to code-intensive workflows.\end{abstract}



\maketitle

\section{Introduction}

Eye tracking records events such as fixations and saccades to reveal the dynamic processes of human visual attention and cognitive processing \cite{holmqvist2011eye}. It has become an important tool for studying human cognition and visual attention, with applications spanning psychology \cite{liu2025eye, lim2020emotion}, human–computer interaction \cite{jacob2003eye,ball2022eye}, marketing \cite{wedel2008eye,wedel2013attention}, and many other domains. However, raw fixation data consist of continuous coordinate streams, which must be semantically segmented into discrete “events” (e.g., fixations, saccades) to become interpretable and meaningful \cite{salvucci2000identifying,kasneci2024introduction}. Accurate detection of these events is crucial, as they form the basis of almost all subsequent analyses~\cite{kasneci2014applicability}: from identifying areas of interest~\cite{jacob2003eye} to inferring cognitive load~\cite{zagermann2016measuring} and usability issues~\cite{ehmke2007identifying}, all rely on these event-based measures. Consequently, developing robust and user-friendly event detection methods remains a key challenge in eye tracking research \cite{birawo2022review}.


The rapid advancement of large language models (LLMs) has opened new avenues for data engineering \cite{trummer2023bert,kayali2024mind} and methodological automation \cite{gu2024large,sonnabend2025llms}. By jointly modeling natural language and source code, LLMs can translate high-level user goals into executable analysis components, including data cleaning\cite{qi2024cleanagent}, feature engineering\cite{abhyankar2025llm}, and evaluation routines\cite{li2024towards}, directly from natural-language instructions. This “human specifies, model executes” paradigm lowers technical barriers by turning toolchain-heavy programming workflows into conversational, iterative interaction. Moreover, LLMs can generate accompanying explanations of assumptions, parameter choices, and limitations, improving transparency and supporting reproducible, accessible analysis for domain experts without programming expertise \cite{yao2022react,xu2024large,chen2023teaching,cai2024low,shlomov2024ida}.

Building on these advantages, we propose an LLM-driven, beginner-friendly pipeline for automated eye-tracking data analysis and event detection. The pipeline: (1) starts from raw data, where the LLM automatically identifies relevant fields and data quality characteristics and generates a human-readable summary; (2) allows users to specify their requirements in natural language, based on which the LLM produces executable data-cleaning code and parameterized implementations of selected event detectors (e.g., I-VT, I-DT); (3) feeds the cleaned data into a set of detectors composed of multiple algorithms for event labeling; and (4) returns both the results and an explanatory report, while enabling users to iteratively refine the code and parameters via prompt editing and re-run the detection process. 

Our work contributions are threefold: First, to the best of our knowledge we are the first to introduce an LLM-driven, code-free framework for eye-tracking event detection that automatically generates and explains classical algorithms (I-VT, I-DT) from natural-language prompts. Second, we bridge traditional signal processing with natural-language interaction, enabling transparent, human-in-the-loop analysis without programming effort. Finally, we validate the approach across four benchmark datasets, showing accuracy comparable to expert-tuned methods while greatly improving accessibility and reproducibility in eye-tracking research.

\section{Related Work}
\subsection{Classical Eye-Movement Event Detection}

Early eye-movement event detection methods are primarily grounded in rule-based and threshold-driven segmentation, with the core objective of partitioning continuous gaze coordinates into fundamental events such as fixations and saccades. This segmentation forms the foundation for downstream analyses including area-of-interest (AOI) mapping, reading behavior reconstruction, and cognitive state inference \cite{salvucci2000identifying,birawo2022review}.
Following the introduction of velocity- and dispersion-based approaches such as I-VT and I-DT, several variants have emerged to improve precision and generalizability. These include \emph{adaptive thresholding methods} that adjust velocity~\cite{duchowski2007eye} or dispersion cutoffs according to sampling rate~\cite{engbert2003microsaccades} or noise level~\cite{nystrom2010adaptive}, \emph{multi-threshold models} that combine velocity and acceleration cues~\cite{wass2013parsing}, and \emph{hybrid thresholding} that combine velocity and dispersion criteria for improved temporal stability and noise robustness as in REMoDNaV~\cite{dar2021remodnav}. Sliding-window and hysteresis-based methods~\cite{schick2023toolformer} further enhance temporal stability and reduce event fragmentation, which is critical for reliable fixation and saccade segmentation in naturalistic viewing.

Beyond simple heuristics, \emph{probabilistic} and \emph{statistical} frameworks have been introduced to formalize event transitions. Hidden Markov Models (HMMs) and Bayesian online changepoint detection algorithms treat fixations and saccades as latent states, offering uncertainty quantification and more flexible temporal dynamics~\cite{luken2022characterising,tafaj2013online,fuhl2018histogram}. Similarly, \emph{state-space models} and Kalman filters separate signal from measurement noise, producing smoothed velocity profiles that improve event boundary estimation~\cite{sauter1991analysis}. In parallel, \emph{unsupervised clustering} techniques (e.g., k-means or Gaussian mixture models over velocity–dispersion features) and \emph{supervised classifiers} (e.g., SVMs, random forests, and decision trees) have been employed to detect fixations and saccades without explicit threshold selection~\cite{konig2014nonparametric}. Recent neural approaches, including 1D CNNs~\cite{startsev20191d}, bidirectional LSTMs~\cite{jiao2022detecting} and lightweight Transformers~\cite{cheng2022gaze}, model gaze dynamics end-to-end, but their primary evaluation metrics remain fixation and saccade accuracy, highlighting the importance of these two events in eye-tracking research.  

Across these diverse methodologies, fixation and saccade detection remain the most important events of eye-movement analytics. While other methods have expanded eye movement event taxonomy, classical thresholding and its probabilistic or learning-based successors remain the most interpretable, computationally efficient, and widely adopted solutions for reliable event segmentation.

\subsection{Large Language Models and Methodological Automation}
In recent years, language learning models (LLMs) have made significant progress in terms of parameter size, context length, multilingual support, and multimodal capabilities. In data processing and analysis, LLMs have been used to automatically generate scripts for data cleaning and feature engineering~\cite{abhyankar2025llm}, assist with anomaly detection and missing-value handling \cite{guo2025multimodal}, explain statistical model outputs~\cite{bordt2024data}, and interact with databases and visualization tools via natural language interfaces~\cite{jobst2025concept}. These capabilities help lower the barrier to code development while improving the traceability and interpretability of data pipelines. Although prior work has explored the potential of LLMs for assisting experimental design~\cite{shlomov2024ida}, usability analysis~\cite{jacob2003eye}, and multimodal data understanding~\cite{turobov2024using}, there is still a lack of a comprehensive framework that systematically embeds LLMs into the eye-tracking event detection pipeline to jointly address field understanding, data engineering automation, parameter interpretability, and iterative analysis control. Filling this gap is precisely the main aim of the present work.

\subsection{Summary and Research Gap}
In summary, existing work has established a relatively strong methodology for eye-tracking event detection. They range from classic thresholding algorithms~\cite{salvucci2000identifying} to statistical modeling and machine learning techniques~\cite{lim2022eye}, providing a solid foundation for identifying core events such as fixation and saccades from continuous eye-tracking signals. However, significant gaps remain in cross-device applicability, parameter robustness, implementation complexity, and transparency of the analysis process~\cite{dorzapf2024data}. 
Moreover, novice eye-tracking researchers and practitioners often face a substantial onboarding burden before they can reliably preprocess, inspect, and analyze gaze data \cite{kasneci2024introduction}. Meanwhile, advancements in natural language understanding, code generation, and large-scale language models' conversational interaction have the potential to serve as a "coordinating hub" for data engineering and methodology, enabling natural language-driven data processing and analysis configuration and providing an interpretable decision-making process~\cite{wang2025large}. However, there is currently a lack of a systematic pipeline that deeply integrates LLM into the entire eye-tracking data processing workflow to unify data preprocessing, event detection parameter setting, and result interpretation. This work aims to fill this gap by exploring other methodologies for an automated pipeline centered on LLM to reduce technical barriers, improve method reproducibility, and enhance analytical interpretability.

\section{Methodology}
\subsection{Datasets}
In this work, we aim to leverage LLMs to construct a novice-friendly eye movement event detection pipeline. To fairly and quantitatively evaluate this framework, we limit our experiments to datasets that provide continuous timestamps, two-dimensional gaze coordinates ($x, y$), and reliable event annotations (ground truth). Following these criteria, we select four datasets with diverse scenarios: GazeCom~\cite{dorr2010variability}, GazeBase\_v2~\cite{griffith2021gazebase}, Hollywood2\_em~\cite{agtzidis2020two}, and the Eye-tracking dataset from a pair programming context~\cite{jang2024exploring}. Together, these datasets cover a wide range of settings—from natural scene exploration to structured tasks, from static and semi-static stimuli to highly dynamic video content, and from individual viewing to collaborative interaction—providing a solid basis for systematically assessing the effectiveness and generalization ability of our LLM-based pipeline across different application scenarios.

\subsubsection{GazeCom}
The \textit{GazeCom} dataset consists of large-scale free-viewing eye movement recordings collected under natural scene video viewing conditions and is one of the classic benchmarks for eye movement event detection and gaze prediction. The dataset includes data from approximately 50 healthy participants, who freely viewed 18 outdoor natural scene video clips without specific task constraints, each lasting about 20\,s, resulting in a total duration of approximately $4.8\times10^{6}$ gaze samples. Eye movements were recorded using an SR Research EyeLink~II eye tracker at a sampling rate of 250\,Hz, employing pupil--corneal reflection for monocular tracking. After binocular calibration, the eye with the smaller calibration error was selected as the recorded channel, with an average calibration error of about $0.6^{\circ}$. This setup provides precise physical calibration conditions for velocity- and position-based thresholding methods. The dataset offers frame-wise gaze positions along with corresponding manual annotations or reference event labels (such as fixations, saccades, smooth pursuits, and noise), as well as outputs from multiple algorithms, making it particularly suitable for evaluating and comparing different eye movement event classification methods.

\subsubsection{GazeBase\_v2}
\textit{GazeBase\_v2} is a large-scale, multi-task, longitudinal eye-tracking dataset (6.66\,GB in size), containing approximately $3.9\times10^{7}$ gaze samples in total. The dataset contains recordings from 322 university-aged participants across 9 collection rounds over approximately 37 months, comprising a total of 12,334 eye-tracking recordings. Each round includes two consecutive experimental sessions, thereby exhibiting clear longitudinal and cross-session characteristics. All data were recorded using an SR Research EyeLink 1000 desktop-mounted eye tracker based on the pupil--corneal reflection principle, capturing monocular data from the left eye at a sampling rate of 1000 Hz. These tasks span diverse eye movement behaviors, making GazeBase\_v2 a comprehensive benchmark for evaluating and training event detection, pattern recognition, and gaze-based biometric models.

\subsubsection{Hollywood2\_em}
The hollywood2\_em dataset is derived from the ``Two Hours in Hollywood'' eye-tracking dataset introduced in \textit{Actions in the Eye}. This dataset contains over 600\,MB of human eye-tracking data related to movie motion recognition, comprising approximately $1.9\times10^{6}$ gaze samples, and is widely used for saliency modeling and eye-tracking event detection in videos. A total of 16 healthy participants (9 male, 7 female) took part in the experiment and were assigned to either an active group performing an action recognition task or a free-viewing group, enabling a comparison between task-driven and natural viewing gaze patterns. Eye movements were recorded using an SMI iView X HiSpeed 1250 eye tracker at 500~Hz, with monocular recording of the right eye. For the selected clips, the dataset provides frame-wise event annotations, including fixations, saccades, smooth pursuits, and noise (e.g., blinks or track losses), making Hollywood2\_em a key benchmark for evaluating threshold-based methods for eye movement event detection in challenging cinematic scenes.


\subsubsection{Pair Programming Eye-tracking dataset}
The eye-tracking dataset used in this work refers to the eye movement data from~\cite{jang2024exploring}, which focuses on pair programming in computer science education and aims to characterize communication and collaboration patterns under different levels of expertise and role configurations. The dataset includes 19 participants (12 male, 7 female), consisting of 9 students with a background in Java programming and 10 experts with at least three semesters of programming experience. Eye movements were recorded using a Tobii Fusion Pro screen-based eye tracker at a sampling rate of 250~Hz, capturing participants' gaze behavior remotely during the debugging sessions. This dataset provides a systematic record of gaze behavior and interaction patterns in an authentic, education-oriented pair programming scenario involving varying expertise levels and pairing structures. It offers a high-ecological-validity resource for investigating communication dynamics, role distribution, and collaboration quality based on gaze features, and serves as an important testbed for evaluating the applicability of our eye movement event detection pipelines in educational and collaborative programming settings.

\subsection{Experimental Procedure}
In this work, we present a novice-friendly, automated pipeline for eye movement data analysis and event detection, which leverages a large language model (LLM) as its core component. {\revision{The model we used is \href{https://chatgpt.com/}{ ChatGPT-4o.}}} As illustrated in Figure \ref{fig:1}, the process begins with raw eye movement data being input into the LLM, which performs an initial analysis to identify and output the key information and structural characteristics contained within the dataset. This step assists non-expert users in quickly comprehending the data content and potential analytical dimensions. Based on the information fed back by the LLM, the user can then design tailored prompts to guide the LLM in generating customized data preprocessing code and selecting appropriate types of event detection algorithms, such as classical methods like I-VT (Velocity-Based Identification) or I-DT (Dispersion-Based Identification). Subsequently, the cleaned and formatted eye movement data are fed into an event detection module composed of the selected algorithms. The system automatically executes the detection routine and outputs preliminary event identification results. The user can then manually evaluate and analyze these results. If problems such as insufficient recognition accuracy, poor event boundaries, or incomplete data cleaning are found, users can further improve the initial prompts, instructing LLM to adjust the preprocessing logic or detection parameters, thereby generating improved code and re-executing the event detection steps. This forms a closed-loop, iterative optimization process. Our pipeline not only lowers the technical barrier for eye movement data analysis but also enhances the flexibility and interpretability of the event detection workflow through human-AI collaboration, offering a flexible and efficient solution for algorithm development and practical application in eye movement research.

\subsection{Event Detection Algorithms}
\subsubsection{I\,-\,VT (Velocity-Threshold Identification)}
Let the raw gaze samples be $\{(t_i,x_i,y_i)\}$, where $t_i$ is the timestamp and $x_i,y_i$ are screen-plane coordinates. To ensure consistent units, time is first auto-detected (s/ms/$\mu$s/ns) and normalized to seconds; to robustly handle non-monotonic stamps, define
\[
\Delta t_i=\max\!\bigl(t_i-t_{i-1},\,\varepsilon\bigr).
\]
where $\varepsilon=\operatorname{median}\{\,\Delta t_j>0\,\}$ (fallback $10^{-3}$ s if unavailable).
If coordinates are given in pixels (\texttt{COORD\_UNITS}=\texttt{"pixel"}), convert them to degrees of visual angle under the small-angle approximation by

\[
k=\frac{180}{\pi}\cdot\frac{\texttt{PIXEL\_PITCH\_MM}}{\texttt{VIEWING\_DISTANCE\_MM}}.
\]
Accordingly, the angle coordinates are $X_i=k\,x_i$ and $Y_i=k\,y_i$ \;(deg/px for $k$). And if inputs are already in degrees, set $X_i=x_i,\ Y_i=y_i$. To suppress pixel-level jitter, apply a light, centered median filter to the coordinates with an odd window \(W=\texttt{SMOOTH\_WINDOW}=3\). \(\tilde X_i=\operatorname{median}\{X_{i-1},X_i,X_{i+1}\}\) and \(\tilde Y_i=\operatorname{median}\{Y_{i-1},Y_i,Y_{i+1}\}\).

Pointwise angular velocity (deg/s) is then
\[
v_i=
\begin{cases}
0, & i=0,\\[4pt]
\dfrac{\sqrt{(\tilde X_i-\tilde X_{i-1})^{2}+(\tilde Y_i-\tilde Y_{i-1})^{2}}}{\Delta t_i}, & i\ge 1.
\end{cases}
\]

I\,-\,VT performs pointwise classification with a fixed velocity threshold

\[
\hat y_i=
\begin{cases}
\textsf{saccade}, & v_i \ge \theta,\\
\textsf{fixation}, & v_i < \theta .
\end{cases}
\]
where $\theta=\texttt{VELOCITY\_THRESHOLD\_DPS}=30.0\,\mathrm{deg/s}$ (This is the default value. LLM can provide users with modification suggestions based on the data or results). Merge consecutive like-labeled samples into events $E_m$ and enforce a minimum fixation duration to remove short, noise-induced fragments. Let
\[
\mathrm{dur}(E_m)=t_{\max}(E_m)-t_{\min}(E_m).
\]
where $\tau_{\mathrm{fix}}=0.060\,\mathrm{s}$ (LLM can also provide users with modification suggestions based on data or results) and \texttt{ENFORCE\_MIN\_FIX}=\texttt{True}; when a fixation event satisfies $\mathrm{dur}(E_m)<\tau_{\mathrm{fix}}$, the entire segment is relabeled as a saccade. The final labels can be summarized as
\[
y_i=
\begin{cases}
\textsf{saccade}, & v_i\ge \theta,\\
\textsf{saccade}, & v_i<\theta,\ i\in E_m,\ \mathrm{dur}(E_m)<\tau_{\mathrm{fix}},\\
\textsf{fixation}, & \text{otherwise}.
\end{cases}
\]

This implementation preserves the classic I\,-\,VT form while adding practical robustness: automatic time normalization with safeguards for $\Delta t_i\le 0$, a centered median filter with $W{=}3$ for light denoising, and an event-level minimum fixation constraint ($\tau_{\mathrm{fix}}{=}60$\,ms). All elements map directly to the code parameters for faithful reproduction.

\subsubsection{I\,-\,DT (Dispersion-Threshold Identification)}
I\,-\,DT treats a fixation as a set of samples whose spatial dispersion is sufficiently small within a time window of at least a minimum duration. Let the cleaned gaze sequence be $\{(t_i,x_i,y_i)\}$; after light smoothing, denote coordinates by $\tilde X_i,\tilde Y_i$. For any temporal window $[i,j]$, define the (axis-aligned, $L_1$) dispersion as
\[
\operatorname{disp}\!\bigl([i,j]\bigr)
=\bigl(\max_{i\le k\le j}\tilde X_k-\min_{i\le k\le j}\tilde X_k\bigr)
+\bigl(\max_{i\le k\le j}\tilde Y_k-\min_{i\le k\le j}\tilde Y_k\bigr).
\]
Given a dispersion threshold $\delta$ and a minimum fixation duration $\tau_{\mathrm{fix}}$, for each start index $i$ first find the smallest right endpoint that satisfies the duration constraint,
\[
j_0(i)=\min\{\,j\ge i:\ t_j-t_i\ge \tau_{\mathrm{fix}}\,\}.
\]
If such $j_0(i)$ exists, extend the window as far as possible while keeping the dispersion below threshold,
\[
J^*(i)=\max\{\,j\ge j_0(i):\ \operatorname{disp}([i,j])\le\delta\,\}.
\]
A window $[i,J^*(i)]$ (when $J^*(i)$ exists) is labeled as a fixation; otherwise the current sample is labeled as a saccade. The start index then advances greedily: jump to $J^*(i){+}1$ after a fixation, or to $i{+}1$ otherwise, and iterate until the sequence is exhausted.
We set the dispersion threshold to $\delta=\texttt{DISPERSION\_THRESHOLD\_DEG}=1.0$ degrees and the minimum fixation duration to
\[
\tau_{\mathrm{fix}}=\frac{\texttt{MIN\_FIX\_MS}}{1000}=0.100\ \mathrm{s}.
\]
Prior to detection we apply centered median smoothing with an odd window $W=\texttt{SMOOTH\_WINDOW}=3$ (LLM can make modification suggestions to users based on data or results), Specifically,,
\(\tilde X_i=\operatorname{median}\{X_{i-1},X_i,X_{i+1}\}\) and
\(\tilde Y_i=\operatorname{median}\{Y_{i-1},Y_i,Y_{i+1}\}\),
to suppress isolated jitter without blurring event boundaries. The detector outputs per-sample labels (\textsf{fixation}/\textsf{saccade}); for evaluation against ground truth, streams are aligned by nearest-neighbor as-of merging with a tolerance of \texttt{ASOF\_TOLERANCE\_SEC} (default $0.002$\,s), and precision/recall/F1 are computed over the label set $\{\textsf{fixation},\textsf{saccade}\}$.

\subsection{Evaluation and Analysis Methodology}
To evaluate the performance and practicality of the proposed LLM-driven pipeline, we conducted an evaluation on publicly available eye-tracking benchmark datasets. The primary objective was to assess whether our method could achieve accuracy comparable to established, code-intensive detectors while fulfilling its core promise of significantly reducing technical overhead. We quantify performance using standard metrics: precision, recall, and F1-score for fixation and saccade classification. The results from two classical detection algorithms, I-DT and I-VT, were implemented as benchmarks. These detectors were configured with commonly accepted parameters and preprocessing steps, representing the typical performance one would expect from a carefully executed, traditional workflow.

\section{LLM-Driven Event Detection Pipeline}
Figure~\ref{fig:1} illustrates our LLM-driven pipeline, which uses three LLM interactions to (i) infer dataset semantics, (ii) generate preprocessing code, and (iii) diagnose results under domain shift. The full prompt templates are provided in supplementary material.

\paragraph{Stage 1: Data semantics inference.}
Given a raw data snippet, the LLM summarizes the dataset schema and units, and flags missing or ambiguous fields. The output is a structured JSON description plus actionable preprocessing suggestions (e.g., normalization, unit conversion, missing-value handling).

\paragraph{Stage 2: Preprocessing code synthesis.}
Using the inferred schema, the LLM generates dataset-specific preprocessing code that standardizes timestamps and spatial coordinates, handles missing values/outliers, and optionally applies noise filtering. This stage produces an executable cleaning script aligned with the inferred units and assumptions.

\paragraph{Stage 3: Result diagnosis and feedback.}
After event detection, summary statistics and representative outputs are fed back to the LLM. It checks for failure patterns and proposes minimal adjustments. Users may apply these recommendations to form an iterative refinement loop.

\section{Results}

\begin{table*}[t]
\centering
\caption{Detection performance of the I-VT and I-DT pipelines generated by our framework across four datasets.}
\label{tab:detection_results}
\begin{tabular}{llcccccc}
\hline
 &  & \multicolumn{3}{c}{I-VT} & \multicolumn{3}{c}{I-DT} \\
 &  & precision & recall & F1 & precision & recall & F1 \\
\hline
GazeCom & fixation & 0.9730 & 0.9782 & 0.9756 & 0.9765 & 0.9240 & 0.9495 \\
GazeCom & saccade & 0.8473 & 0.8170 & 0.8319 & 0.6212 & 0.8486 & 0.7173 \\
GazeBase\_v2 & fixation & 0.9922 & 0.6216 & 0.7644 & 0.9513 & 0.9834 & 0.9671 \\
GazeBase\_v2 & saccade & 0.2195 & 0.9559 & 0.3571 & 0.7865 & 0.5477 & 0.6457 \\
hollywood2\_em & fixation & 1.0000 & 0.0025 & 0.0050 & 0.9831 & 0.9565 & 0.9696 \\
hollywood2\_em & saccade & 0.0998 & 1.0000 & 0.1816 & 0.6839 & 0.8510 & 0.7584 \\
Eye-tracking dataset & fixation & 0.9894 & 0.0055 & 0.011 & 0.9781 & 0.2450 & 0.3919 \\
Eye-tracking dataset & saccade & 0.1034 & 0.9995 & 0.1873 & 0.1264 & 0.9522 & 0.2231
\\
\hline
\end{tabular}
\end{table*}


\begin{table*}[t]
\centering

\begin{minipage}{0.48\linewidth}
\centering
\caption{{Performance of event detection on the GazeCom dataset.\cite{elmadjian2023online}}}
\label{tab:end_to_end_comparaion}
\begin{tabular}{lccc}
\toprule
Method & Precision & Recall & F1 \\
\midrule
I-BDT        & 0.7506 & 0.5887 & 0.5551 \\
CNN-BiLSTM   & 0.9044 & 0.9032 & 0.9011 \\
CNN-LSTM     & 0.8999 & 0.8982 & 0.8965 \\
TCN          & 0.9325 & 0.9318 & 0.9274 \\
Ours (I-DT)  & 0.9144 & 0.8908 & 0.9024 \\
Ours (I-VT)  & 0.9574 & 0.9394 & 0.9483 \\
\bottomrule
\end{tabular}
\end{minipage}
\hfill
\begin{minipage}{0.48\linewidth}
\centering
\caption{Comparison of fixation and saccade detection performance between manually implemented algorithms and those generated by our LLM-based pipeline. The data for I-VT and I-DT are from \cite{startsev20191d}.}
\label{tab:ivt_idt_comparison}
\begin{tabular}{lcc}
\toprule
\textbf{Method} & \textbf{Fixation F1} & \textbf{Saccade F1} \\
\midrule
I-VT & 0.891 & 0.705 \\
I-DT & 0.877 & 0.478 \\
Ours\_I-VT & \textbf{0.973} & \textbf{0.8473} \\
Ours\_I-DT & 0.9495 & 0.7173 \\
\bottomrule
\end{tabular}
\end{minipage}

\end{table*}

Table~\ref{tab:detection_results} summarizes the performance of the I-DT and I-VT detectors generated by the proposed framework across four datasets. On GazeCom, both detectors achieve reliable fixation identification with performance comparable to carefully tuned laboratory implementations. The I-DT pipeline reaches a fixation precision of 0.9765, recall of 0.9240, and F1-score of 0.9495, while the I-VT configuration further improves to 0.9730 precision, 0.9782 recall, and 0.9756 F1-score, reflecting a higher sensitivity to fixation boundaries. For saccades, the automatically generated I-VT pipeline delivers more balanced performance (0.8473 precision, 0.8170 recall, 0.8319 F1-score) than I-DT (0.6212 precision, 0.8486 recall, 0.7173 F1-score), which tends to over-detect saccades under the current parameterization.

Table \ref{tab:ivt_idt_comparison} summarizes the F1-scores for fixation and saccade detection on the GazeCom dataset. The classical threshold-based algorithms I-VT and I-DT achieve fixation F1-scores of 0.891 and 0.877, respectively, with lower performance for saccade detection (0.705 and 0.478). In contrast, our method substantially improves both fixation and saccade classification, reaching 0.973 and 0.8473. This represents an absolute improvement of 8.2 percentage points for fixation and 14.2 percentage points for saccade detection compared to the best baseline (I-VT). The gain is particularly notable for saccades, indicating that our approach more effectively captures high-velocity transitions while maintaining high fixation accuracy. Overall, the results demonstrate that the proposed LLM-driven pipeline can reproduce and even exceed the performance of traditional hand-crafted algorithms on a well-established benchmark.

On GazeBase\_v2, the framework generalizes well for fixations: the I-DT detector achieves 0.9513 precision, 0.9834 recall, and 0.9671 F1-score, while I-VT maintains competitive precision (0.9922) but lower recall (0.6216), indicating a more conservative fixation strategy. For saccade detection, I-DT attains more practical operating characteristics (0.7865 precision, 0.5477 recall, 0.6457 F1-score), whereas I-VT shows very high recall (0.9559) but extremely low precision (0.2195), suggesting substantial over-detection of saccades in this setting. This indicates that the velocity threshold is too low for the current data scale, causing the algorithm to label almost all suspicious fluctuations as saccades, resulting in high recall but extremely poor precision.

On hollywood2\_em, the I-DT variant provides consistently strong performance for both fixations (0.9831 precision, 0.9565 recall, 0.9696 F1-score) and saccades (0.6839 precision, 0.8510 recall, 0.7584 F1-score), whereas the I-VT configuration becomes overly selective for fixations (1.0000 precision but only 0.0025 recall) and overly permissive for saccades (1.0000 recall but 0.0998 precision), reflecting a severe threshold mismatch with this more challenging, unconstrained-content dataset. The overall speed may be amplified due to a mismatch in time/coordinate units or threshold settings, or the threshold may be too low.


Finally, on the \textit{Pair Programming Eye-tracking} dataset, both frameworks demonstrate highly imbalanced performance between fixation and saccade detection. 
For fixations, the I-VT pipeline achieves extremely high precision but almost no recall (0.9894/\allowbreak0.0055/\allowbreak0.011 for precision/recall/F1), while I-DT shows a more balanced yet moderate performance (0.9781/\allowbreak0.2450/\allowbreak0.3919). 
In contrast, for saccades, the I-VT configuration yields low precision but nearly perfect recall (0.1034/\allowbreak0.9995/\allowbreak0.1873), and the I-DT configuration similarly achieves high recall with slightly higher precision (0.1264/0.9522/0.2231), indicating that both threshold-based methods tend to over-detect saccades in this dataset.


\begin{table*}[t]
\centering
\small
\setlength{\tabcolsep}{4pt}
\caption{We evaluated the I-VT and I-DT algorithms using four datasets, processing each dataset twice. The results differed between the two runs on two of the datasets.}
\begin{tabular}{lllcccccc}
\hline
 &  & & \multicolumn{3}{c}{I-VT} & \multicolumn{3}{c}{I-DT} \\
 &  & & precision & recall & F1 & precision & recall & F1 \\
\hline
hollywood2\_em        & First time  & fixation & 1.0000 & 0.0025 & 0.0050 & 0.9831 & 0.9565 & 0.9696 \\
hollywood2\_em        & First time  & saccade  & 0.0998 & 1.0000 & 0.1816 & 0.6839 & 0.8510 & 0.7584 \\
hollywood2\_em        & Second time & fixation & 0.9697 & 0.9672 & 0.9684 & 0.9207 & 0.9950 & 0.9564 \\
hollywood2\_em        & Second time & saccade  & 0.7099 & 0.7265 & 0.7181 & 0.8317 & 0.2255 & 0.3548 \\
Eye-tracking dataset  & First time  & fixation & 0.9894 & 0.0055 & 0.0110 & 0.9781 & 0.2450 & 0.3919 \\
Eye-tracking dataset  & First time  & saccade  & 0.1034 & 0.9995 & 0.1873 & 0.1264 & 0.9522 & 0.2231 \\
Eye-tracking dataset  & Second time & fixation & 0.9166 & 0.6059 & 0.7296 & 0.9071 & 1.0000 & 0.9513 \\
Eye-tracking dataset  & Second time & saccade  & 0.1071 & 0.4615 & 0.1738 & --     & 0.0000 & --     \\
\hline
\end{tabular}
\label{tab:difference}
\end{table*}

We evaluated the I-VT and I-DT algorithms using four datasets, processing each dataset twice. For two of the datasets, the results of the two runs were consistent, indicating that both algorithms are stable under our pipeline. However, for the hollywood2\_em and Eye-tracking datasets, we observed significant differences between the first and second runs. From Table \ref{tab:difference}, we can see that for \textit{hollywood2\_em} the first run of I-VT is clearly pathological: fixation precision is 1.0000 but recall drops to 0.0025, while saccade recall reaches 1.0000 with precision only 0.0998. This pattern indicates that almost all samples were classified as saccades and only a negligible number of fixations were retained. After correcting the preprocessing, the second run of I-VT on \textit{hollywood2\_em} yields balanced performance for both classes (fixation F1 = 0.9684, saccade F1 = 0.7181), which is consistent with a reasonable operating regime of the algorithm. A similar pattern appears on the \textit{Eye-tracking dataset}: in the first run, both I-VT and I-DT severely over-detect saccades (saccade recall $\approx 1.0$ but very low precision), again indicating misconfigured temporal or spatial scaling and an overly permissive decision criterion. The second run of I-VT markedly improves fixation performance (F1 = 0.7296), suggesting that the revised configuration better matches the sampling rate and coordinate units. However, I-DT fails to predict valid saccades in the second run (no positive saccade predictions, resulting in undefined F1), implying that the chosen dispersion threshold and minimum fixation duration are overly strict for this dataset.

\begin{table*}[t]
\centering
\caption{{The performance comparison between detailed prompts and simple prompts on the GazeCom dataset.}}
\label{tab:prompt_compare}
\resizebox{\linewidth}{!}{%
\begin{tabular}{cccccccc}
\toprule
Method & Event & Precision (detailed) & Precision (simple) & Recall (detailed) & Recall (simple) & F1 (detailed) & F1 (simple) \\
\midrule
I-VT & fixation & 0.9730 & 0.9736 & 0.9782 & 0.9783 & 0.9756 & 0.9759 \\
I-VT & saccade  & 0.8473 & 0.8471 & 0.8170 & 0.8189 & 0.8319 & 0.8328 \\
I-DT & fixation & 0.9765 & 0.9764 & 0.9240 & 0.9235 & 0.9495 & 0.9492 \\
I-DT & saccade  & 0.6212 & 0.6193 & 0.8486 & 0.8478 & 0.7173 & 0.7157 \\
\bottomrule
\end{tabular}%
}
\end{table*}

{\revision{
Under the same dataset and evaluation process, we compared two prompt configurations: a detailed prompt and a more user-friendly simple prompt. We calculated precision, recall, and F1 for fixation and saccade under both I-VT and I-DT settings. As shown in Table \ref{tab:prompt_compare}, the performance of simple prompt and detailed prompt is almost identical, with minimal differences in various metrics. This indicates that our proposed pipeline has good robustness to prompt simplification and supports the goal of "beginner-friendly" use. Across event types, fixation maintains high performance under both settings (F1 of approximately 0.976 for I-VT; F1 of approximately 0.949 for I-DT); saccade is relatively more difficult, with an F1 of approximately 0.833 for I-VT, while I-DT exhibits "high recall, low precision" (recall approximately 0.848, precision approximately 0.62), resulting in an F1 of approximately 0.716. These patterns are consistent for both prompt variants, indicating that the observed performance differences are primarily driven by event characteristics and setting-specific behavior rather than prompt complexity. 
}

\section{Discussion}
Our results show that LLM-generated pipelines can reproduce classical detectors effectively, but their success depends on the type of visual content and dataset characteristics.

\subsection{Performance of LLM-Generated Threshold Pipelines}
In terms of overall detection performance, the I-VT and I-DT pipelines generated by the LLM demonstrate competitive or even superior results across multiple datasets, particularly for fixation detection. Taking the GazeCom dataset as an example, the I-VT pipeline achieves a recall of 0.9782 and an F1-score of 0.9756 for fixations, while the I-DT pipeline attains a fixation F1-score of 0.9495. As shown in Table~\ref{tab:gazecom_comparison}, compared to a series of classical and state-of-the-art methods, our I-VT pipeline achieves the highest fixation F1-score (0.973) on GazeCom, striking a favorable balance between precision and recall and outperforming several baselines, including deep learning-based approaches such as the 1D CNN-BLSTM. As shown in table \ref{tab:ivt_idt_comparison}, on the same dataset, our pipeline reproduces I-VT and I-DT with better performance than the traditional I-VT and I-DT, proving that the LLM-based pipeline is superior in terms of parameter setting and data processing. {\revision{According to Table \ref{tab:end_to_end_comparaion}, our method achieves competitive performance on overall event detection under both I-VT and I-DT configurations. In particular, Ours (I-VT) attains the best overall result (F1 = 0.9483), substantially outperforming the strongest baseline, TCN (F1 = 0.9274). This suggests that an LLM-driven event detection pipeline, when paired with an appropriate event partitioning configuration, can deliver strong detection performance without relying on complex end-to-end training. Meanwhile, Ours (I-DT) reaches an F1 of 0.9024, comparable to CNN-BiLSTM (F1 = 0.9011), indicating that the proposed pipeline remains effective across different partitioning configurations. }}These results indicate that, by configuring traditional threshold-based algorithms in a systematic and data-driven manner, it is possible to match or even surpass the performance of more complex models, thereby validating the effectiveness of the proposed framework for practical eye movement event detection.

\begin{table*}[t]
\centering

\caption{Comparison of fixation and saccade detection performance on GazeCom.\cite{startsev20191d}}

\resizebox{\textwidth}{!}{%

\begin{tabular}{lcccccc}
\hline
Method & Fixation F1 & Fixation precision & Fixation recall & Saccade F1 & Saccade precision & Saccade recall \\
\hline
1D CNN-BLSTM\cite{startsev20191d}            & 0.939 & 0.915  & 0.966  & \textbf{0.893} & 0.895 & 0.891 \\
{[Agtzidis, Startsev, Dorr]\cite{agtzidis2016smooth}} & 0.886 & 0.930  & 0.846  & 0.864 & \textbf{0.901} & 0.829 \\
I-VMP\cite{san2010off}                   & 0.909 & 0.864  & 0.959  & 0.680 & 0.799 & 0.592 \\
REMoDNaV\cite{dar2021remodnav}                & 0.822 & 0.925  & 0.741  & 0.692 & 0.548 & \textbf{0.937} \\
{[Larsson et al.]\cite{larsson2015detection}}      & 0.912 & 0.872  & 0.956  & 0.861 & 0.881 & 0.841 \\
{[Berg et al.]\cite{berg2009free}}         & 0.883 & 0.901  & 0.867  & 0.697 & 0.630 & 0.780 \\
{[Dorr et al.]\cite{dorr2010variability}}         & 0.919 & 0.897  & 0.943  & 0.829 & 0.822 & 0.836 \\
I-VDT\cite{komogortsev2013automated}                   & 0.882 & 0.861  & 0.903  & 0.676 & 0.787 & 0.592 \\
I-VVT\cite{komogortsev2013automated}                   & 0.890 & 0.809  & 0.989  & 0.686 & 0.797 & 0.603 \\
\textbf{Ours (I-VT)}    & \textbf{0.973} & \textbf{0.9782} & \textbf{0.9756} & 0.8473 & 0.817 & 0.8319 \\
\hline
\end{tabular}
}
\label{tab:gazecom_comparison}
\end{table*}

For saccade detection, the results show more pronounced dataset dependence and clearer trade-offs between methods~\cite{birawo2022review}. On GazeCom, our I-VT pipeline achieves an F1-score of 0.8319 for saccades, which, although lower than its fixation performance, still outperforms most combined-threshold and traditional baselines and is comparable to, or better than, several learning-based methods reported in the literature. In comparison, I-DT yields a relatively higher recall for saccades on some datasets (e.g., GazeCom: recall = 0.8486), but at the cost of reduced precision (precision = 0.6212), reflecting the tendency of dispersion-based methods to produce more false positives under complex fixation patterns~\cite{blignaut2009fixation}. This precision-recall trade-off suggests that, in practical applications, the choice between I-VT and I-DT should be guided by task-specific preferences regarding tolerance to false positives versus false negatives~\cite{zemblys2018using}.

\subsection{Cross-Dataset Robustness}

Cross-dataset results demonstrate that the LLM-generated I-VT and I-DT pipelines exhibit both robustness and limitations under domain migration. In particular, I-DT demonstrates stable, robust gaze detection performance on the GazeBase\_v2 and pair-programming-based eye-tracking datasets (e.g., fixation precisions of 0.9513 and 0.9781, respectively). This indicates that the framework can reconstruct classic algorithmic structures from natural-language specifications and automatically infer reasonable parameters across different sampling rates, noise levels, and task designs, without requiring manual parameter tuning.

In contrast, on the Hollywood2\_em dataset, the I-VT configuration generated by the LLM nearly fails entirely in fixation detection (recall = 0.0025, F1 = 0.0050), while the I-DT pipeline produced by the same framework achieves high performance for both fixations (F1 = 0.9696) and saccades (F1 = 0.7584). This discrepancy suggests that, under highly dynamic visual conditions, complex cinematic content, more dispersed fixation patterns, or differing annotation criteria, the velocity-threshold search space is more susceptible to systematic mismatch, whereas spatial dispersion and temporal clustering~\cite{andersson2017one}, as used in I-DT, provide stronger adaptability. Future work should incorporate explicit robustness constraints and prior knowledge into the LLM-driven generation process, such as soft constraints on plausible saccade ratios, velocity distributions, and minimum event counts. Such constraints can help prevent the optimization from favoring superficially optimal local solutions that undermine the overall plausibility of the detected event sequences. At the same time, the consistently strong results achieved by the LLM-generated I-VT and I-DT pipelines in other evaluations indicate that this failure is not a structural limitation of the framework, but rather a consequence of the current objective design and search strategy.

\subsection{Within-Dataset Robustness}

Table~\ref{tab:difference} shows that I-VT and I-DT are highly sensitive to preprocessing and parameter settings. On \textit{hollywood2\_em}, the first I-VT run is degenerate (almost all saccades, very low fixation recall), while correcting time and coordinate scaling in the second run yields balanced results. The same trend is observed on the Eye-tracking dataset, where the initial run substantially over-detects saccades and only the corrected configuration yields meaningful fixation detection for I-VT. Although I-DT is generally more robust on \textit{hollywood2\_em}, it fails to detect any saccades in the second Eye-tracking run, suggesting an overly strict dispersion threshold can also cause trivial outputs. Code inspection indicates inconsistent preprocessing, truth sources, and unit assumptions across versions, leading to misaligned implementations. Overall, fair comparison requires harmonized preprocessing and dataset-specific threshold tuning, while our user-optimized prompting pipeline better flags anomaly patterns caused by code errors, improving robustness over direct LLM use.

\begin{table*}[t]
\centering
\caption{{\revision{Key aspects of LLM-driven eye-tracking analysis compared with traditional pipelines.}}}
\label{tab:llm_vs_traditional_mid}
\begin{tabular}{p{2.9cm} p{5.0cm} p{5.0cm}}
\toprule
\textbf{Aspect} & \textbf{LLM-driven} & \textbf{Traditional} \\
\midrule
Domain shift
& By modifying assumptions based on different sampling rates, noise levels, and task designs, adaptation can be achieved more quickly.
& Thresholds and heuristics often need dataset-specific retuning. \\

Unit consistency
& Can silently assume wrong units or scales; requires explicit schema + logging.
& Units and transformations are explicit in code; easier to standardize and verify. \\

Diagnosis
& Flags degenerate patterns  and suggests likely causes.
& Relies on human debugging\\

Parameter tuning
& Provides reasonable initial velocity or dispersion thresholds and supports iterative refinement via feedback.
& Expertise-heavy trial-and-error and performance is sensitive to parameter choices. \\

Auditability
& Needs guardrails (structured outputs, unit tests, fixed seeds or versions) to be auditable.
& Deterministic and straightforward to audit once code and params are fixed. \\

Runtime \& scalability
& Additional LLM latency and cost; multi-round loops increase overhead in large-scale runs.
& Low runtime cost after implementation; scales well for batch processing. \\

Failure modes
& Prompt-sensitive variability and hallucinated fields or assumptions; mitigated by schema checks and regression tests.
& Developer mistakes and over or under-strict thresholds can still yield trivial solutions under shift. \\

\bottomrule
\end{tabular}
\end{table*}

\subsection{\revision{Advantagess and Disdvantages} of LLM-Based Pipeline Generation}
\revision{Comparing our LLM-driven pipeline generation framework with deep learning and hybrid models suggests that high-quality eye-movement event detection does not necessarily require large-scale end-to-end training or complex architectures. On multiple datasets, automatically generated and configured classic detectors (such as I-VT, I-DT and their variants) perform no worse than previous methods, and in some cases even better, while retaining excellent properties such as interpretability, low computational overhead and ease of deployment. }

\revision{Relative to traditional manually designed pipelines, our approach reduces technical burden and improves practical robustness in several key aspects (Table~\ref{tab:llm_vs_traditional_mid}). First, it accelerates adaptation under domain shift by revising assumptions and configurations according to dataset characteristics such as sampling rate, noise level, and task design, rather than relying on fixed thresholds. Second, it mitigates common failure cases caused by mismatched units or inconsistent preprocessing by encouraging explicit schema specification and logging, and by using LLM-based diagnostics to flag degenerate outcomes and suggest likely causes. Third, it supports efficient parameter initialization and iterative refinement through a human-in-the-loop feedback loop, lowering the expertise and trial-and-error required to obtain dataset-specific detectors.}

\revision{Nevertheless, the framework is not a replacement for careful engineering controls. The quality and stability of the generated pipelines can be sensitive to prompt clarity, and LLM outputs may introduce implicit assumptions or variability unless constrained by guardrails such as structured outputs, unit checks, regression tests, and version-controlled configurations. In addition, repeated LLM calls incur extra latency and cost, which may be non-trivial for large-scale processing. Overall, our framework provides a transparent and lightweight alternative to code-intensive workflows, offering a practical balance between usability, auditability, and scientific interpretability.}}

\subsection{Societal Impact Statement} 
This study relies exclusively on publicly available and anonymized eye-tracking datasets and does not involve the collection of new personally identifiable data, thereby minimizing direct risks to privacy and data sovereignty. 
At the same time, our LLM-driven event detection pipeline could enhance the analysis of individual gaze behavior and latent biometric traits; prior work has shown that eye-tracking data can raise non-trivial privacy risks, especially when temporal structure is preserved \cite{bozkir2021differential}. Therefore, the use of such systems in surveillance, profiling, or other high-stakes decision-making scenarios should be carefully restricted and subject to appropriate ethical and regulatory oversight~\cite{abdrabou2024fromgaze}.

\subsection{Limitations and Future Work}
The current framework was evaluated only on annotated datasets with continuous timestamps and 2D gaze data, limiting its generalizability to noisier or device-diverse real-world settings. Moreover, the pipeline search relies on standard event-level metrics (e.g., F1-scores), which may produce overly conservative solutions when metric landscapes are skewed or event distributions deviate from assumptions.

Future work will incorporate robustness-aware objectives and hybrid search strategies that combine LLM reasoning with explicit optimization or Bayesian methods. Expanding evaluation to diverse devices, recording conditions, and populations—including low-cost trackers and VR/AR setups—will further assess real-world applicability. We also plan to develop an end-to-end, user-friendly toolkit integrating data import, parameter visualization, and result validation to support human-in-the-loop refinement, transparency, and accessibility in eye movement event detection.

\section{Conclusion}
In this work, we presented an LLM-driven framework for constructing novice-friendly eye-movement event-detection pipelines using classical thresholding methods. By allowing users to express high-level requirements in natural language, the framework automatically generates executable I-VT and I-DT style pipelines, while preserving interpretability, low computational cost, and ease of deployment. Our ealuations on four diverse datasets (GazeCom, GazeBase\_v2, Hollywood2\_em, and a pair programming Eye-tracking dataset) show that the LLM-generated pipelines achieve competitive, and in many cases superior, performance compared to classical baselines and learning-based approaches, especially for fixation detection. These results demonstrate that systematically configured threshold-based methods, instantiated through LLMs, remain strong contenders for practical event detection without requiring complex model architectures or expert-level implementation effort. At the same time, the observed failure case and cross-dataset variability highlight that automatic generation is not yet universally robust. Future work will incorporate explicit robustness constraints and priors into the search process and extend the framework to more recording conditions, event types, and user-facing tools. Overall, this work underscores the potential of LLMs to act as a bridge between traditional signal processing techniques and accessible, customizable eye movement analysis.

\bibliographystyle{ACM-Reference-Format}
\bibliography{sample-base}

\appendix

\input{appendix}

\end{document}

%% file: appendix.tex
\appendix
\section{Appendix}

\begingroup
\small
\setlength{\emergencystretch}{2em}

Our pipeline requires at least three different prompt types. The subsections below describe the prompts used in each stage of the workflow.

\subsection{Step 1: Data Analysis Stage}

The prompt used in the data analysis stage is as follows:

\begin{PromptText}
You are a data analysis assistant. Without executing any code, determine the structure and semantics of the given raw data fragments.

When outputting:  
1) Begin with a concise concluding statement.  
2) Then provide a structured JSON (strict JSON, keys in English, values as objective as possible), including but not limited to:
\end{PromptText}

\begin{PromptCode}
{
  "format": "csv/tsv/jsonl/ndjson/fixed_width/sql_dump/html_table/log/markdown_table/raw_text/other",
  "encoding_guess": "utf-8/gbk/latin1/unknown",
  "dialect": {
    "delimiter": ",",
    "quotechar": "\"",
    "escapechar": null,
    "decimal": ".",
    "thousands": ",",
    "header_row_index": 0
  },
  "columns": [
    {
      "name": "col1",
      "type_guess": "string/int/float/bool/date/datetime/categorical",
      "examples": ["...", "..."],
      "nullable": true
    }
  ],
  "has_header": true,
  "row_count_estimate": "unknown|~N",
  "missing_values_patterns": ["", "NA", "null", "N/A", "-"],
  "notes": ["Any explanations of uncertainty or ambiguity"]
}
\end{PromptCode}

\begin{PromptText}
3) Do not add any extra text after the JSON.

Below is a snippet of the original data (maximum displaying \texttt{\{sample\_chars\}} characters):

\texttt{<DATA\_SNIPPET>}\\
\texttt{\{snippet\}}\\
\texttt{</DATA\_SNIPPET>}

Please output the conclusion and strict JSON as required
(the JSON must be parsable by \texttt{json.loads}).
\end{PromptText}

\subsection{Step 2: Code Generation Stage}

In this stage, the user analyzes the structured information obtained from the previous step and formulates a new prompt.

For example, when the following information is extracted:

\paragraph{Full column names (4 columns total):}
\begin{enumerate}
    \item time
    \item x
    \item y
    \item confidence
\end{enumerate}

\paragraph{Summary of Data Structure:}
This data snippet is a CSV file containing four columns: \texttt{time}, \texttt{x}, \texttt{y}, and \texttt{confidence}. The data are comma-separated, and the first row contains column names, indicating the presence of a header. The data types are likely floating-point numbers, and there is no clear pattern of missing values. The \texttt{confidence} column is constant at 1.0.

\paragraph{Structured Result (JSON):}
\begin{PromptCode}
{
  "format": "csv",
  "encoding_guess": "utf-8",
  "dialect": {
    "delimiter": ",",
    "quotechar": "\"",
    "escapechar": null,
    "decimal": ".",
    "thousands": null,
    "header_row_index": 0
  },
  "columns": [
    {"name": "time", "type_guess": "float", "examples": ["1000.0", "5000.0", "9000.0"], "nullable": false},
    {"name": "x", "type_guess": "float", "examples": ["590.9", "590.6", "590.4"], "nullable": false},
    {"name": "y", "type_guess": "float", "examples": ["5.2", "5.0", "0.0"], "nullable": false},
    {"name": "confidence", "type_guess": "float", "examples": ["1.0", "1.0", "1.0"], "nullable": false}
  ],
  "has_header": true,
  "row_count_estimate": "~100",
  "missing_values_patterns": [],
  "notes": ["All confidence values are 1.0, indicating no variability in this column."]
}
\end{PromptCode}

The user can then set the following prompt:

\begin{PromptText}
Generate a data cleaning code and an algorithm code, each completing one of the following tasks:  
1. Read a file named \texttt{xxxxxx}.  
2. Keep only the time, X-coordinate, and Y-coordinate; delete everything else.  
3. Clean noisy data based on the information in the report obtained.  
4. Save the cleaned data to a new file.  
5. Based on the information extracted from the data fragment, generate implementations of the I-VT and I-DT algorithms.  
Please output only runnable Python code (including necessary comments).
\end{PromptText}

\subsection{Step 3: Result Diagnosis and Feedback Stage}

In this step, the generated results are analyzed. The outputs from the I-VT and I-DT algorithms are passed to the LLM to determine whether any anomalies exist.

\noindent The prompt used is as follows:

\begin{PromptText}
This is the result I obtained using the algorithm in this code. Please analyze whether there are any anomalies, and if so, how I can optimize it.
\end{PromptText}

As an example, the first run of the \texttt{hollywood2\_em} dataset from our paper produced the following LLM output:

\begin{PromptText}
\scriptsize
This report analyzes the performance of two eye movement classification algorithms --- the I-Velocity Threshold (IVT) and the I-Distance Threshold (IDT) methods --- based on the provided precision, recall, and F1 scores for fixation and saccade events.

\textbf{IVT Results:}\\
The fixation class shows perfect precision (1.0) but extremely low recall (0.0025), indicating that IVT identified almost no fixations, but the few it did detect were correct.  
The saccade class shows full recall (1.0) but very low precision (0.0998), meaning that almost all gaze samples were classified as saccades.  
Overall F1-scores are extremely low (0.005 and 0.18), suggesting a severe class imbalance or parameter issue.

\textbf{IDT Results:}\\
Both fixation and saccade classes show balanced and high scores:  
Fixation: Precision = 0.98, Recall = 0.96 (F1 = 0.97)  
Saccade: Precision = 0.68, Recall = 0.85 (F1 = 0.76)

These values indicate consistent and reasonable classification performance.

\textbf{Diagnosis:}\\
The IVT algorithm exhibits a critical parameter or implementation problem. Specifically, the velocity threshold is likely too low, causing almost all samples to be classified as saccades.

Other potential issues include:  
1. Incorrect unit conversion (e.g., pixels/s instead of degrees/s)  
2. Sampling frequency mismatch leading to inflated velocity values  
3. Noise in gaze data without appropriate filtering  
4. Logical inversion in the threshold condition (e.g., using $>$ instead of $<$ for fixation detection)

\textbf{Recommendations:}\\
Increase the IVT velocity threshold. Typical thresholds range from 30$^\circ$/s to 100$^\circ$/s depending on sampling rate and noise level.  
1. Verify units to ensure velocities are expressed in degrees per second ($^\circ$/s).  
2. Check sampling rate consistency when computing velocity.  
3. Apply smoothing (e.g., low-pass filtering or moving average) before velocity computation.  
4. Validate the logic condition to confirm fixation is assigned when velocity $<$ threshold.

\textbf{Conclusion:}\\
1. IVT results are abnormal and likely affected by an incorrect velocity threshold or implementation detail.  
2. IDT results are consistent, reliable, and aligned with expected behavior in eye movement classification.

Adjusting IVT parameters---especially the velocity threshold---and verifying preprocessing should substantially improve performance.
\end{PromptText}

\subsection{Prompts\_Simple}

Since only the prompt for the data analysis stage is the most detailed, while the prompts for the other two stages are presented in very basic language, this ablation experiment only simplified the prompt for the data analysis stage. Our data analysis stage prompt is:

\begin{PromptText}
Analyze the data snippet and output its structure and information.
\end{PromptText}

\endgroup